\documentclass[%
 reprint,
 amsmath,amssymb,
 aps,
]{revtex4-1}
\usepackage{graphicx}
\usepackage{dcolumn}
\usepackage{bm}

\usepackage[german]{babel}
\makeatletter
\newcommand*{\rom}[1]{\expandafter\@slowromancap\romannumeral #1@}
\makeatother

\begin{document}

\title{Geometric measures of discordlike quantum correlations based on Tsallis relative entropy}

\author{Weijing Li}%
\email{liweijing13@mails.ucas.ac.cn}
\affiliation{%
Institute of Mathematics, Academy of Mathematics and Systems Science,\\
Chinese Academy of Sciences, Beijing, 100190, China
}%
\author{}
\affiliation{
 School of Mathematical Sciences,\\ University of Chinese Academy of Sciences, Beijing, 100049, China
}%
\author{}
\affiliation{
 UTS-AMSS Joint Research Laboratory for Quantum Computation and Quantum Information Processing,\\
 Academy of Mathematics and Systems Science, Chinese Academy of Sciences, Beijing, 100190, China
}%

\date{\today}

\begin{abstract}
In this paper we formulate a kind of new geometric measure of quantum correlations. This new measure is in terms of the quantum Tsallis relative entropy and can be viewed as a one-parameter extension quantum discordlike measure that satisfies all requirements of a good measure of quantum correlations. It is of an elegant analytic expression and contains several existing good quantum correlation measures as special cases.
\end{abstract}

\maketitle
\section{Introduction}
Quantum correlation\cite{henderson2001classical}\cite{genovese2005research} beyond entanglement \cite{horodecki2009quantum} has many important applications in the quantum information theory \cite{bera2017quantum}, so finding physically meaningful and mathematically rigorous quantifiers of the quantum correlations are a long-standing central problem among the community of quantum information science, and various useful measures have been established so far \cite{bera2017quantum}\cite{roga2016geometric}\cite{yu2014quantum}\cite{modi2012classical}\cite{adesso2016measures}. Historically, the discordlike quantum correlation measures can be categorized roughly into two different families, namely, those based on quantum entropy theory, and those based on various distance measures of quantum states (also called the geometric quantum discord measures). For the first category one can see \cite{modi2012classical} and \cite{adesso2016measures} for a detailed overview. And the recent paper \cite{hu2018quantum} provided an elegant review of the second category of discordlike quantum correlation measures.

 It has been shown that there are quantum correlations that may arise without entanglement, such as quantum discord \cite{ollivier2001quantum} and measurement-induced nonlocality \cite{luo2011measurement}. For recent progresses and its applications one can see \cite{modi2012classical} \cite{adesso2016measures} .

Quantum coherence \cite{baumgratz2014quantifying}, on the other hand, is another important resource in the quantum mechanics \cite{streltsov2017colloquium}. And it turns out that there is an intimate relationship between the quantum coherence and quantum correlations \cite{hu2018quantum}. Recently, the coherence measure based on quantum Tsallis relative entropy \cite{rastegin2016quantum} had been formulated and its modified version \cite{zhao2018coherence} is showed to be a bona fide coherence quantifier. So it is naturally to ask whether there is or not a similar correlation measure. In this paper, we study the geometric measure of discordlike quantum correlations in terms of quantum Tsallis relative entropy and formulate a class of new definitions of quantum correlation measure which satisfies all the requirements of a good quantum correlation measure. Furthermore,  this kind of measures enjoy an elegant analytic expression. Moreover, our new measure can be viewed as a one-parameter extension of von Neumann relative entropy quantum correlation measure and contains the geometric discord measures based on Hellinger distance \cite{roga2016geometric} and skew information \cite{yu2014quantum} as special cases.

\section{Tsallis relative entropy}
For the convenience of the readers, we simply review some basic properties about Tsallis relative entropy. For two probability distributions $p$ and $q$ on index set $I$, and for $0 < \alpha \ne 1$, the Tsallis relative $\alpha$ entropy is defined as \cite{borland1998information}\cite{furuichi2004fundamental}
\begin{align*}
    D_{\alpha}(p||q)=\frac{1}{\alpha-1}\left(\sum_{j \in I} p_j^{\alpha}q_j^{1-\alpha}-1\right).
\end{align*}

For $0 < \alpha \ne 1$ and real $\xi >0$ the $\alpha$-logarithm is defined as \cite{borland1998information}:
\begin{align*}
    \log_{\alpha}(\xi)=\frac{\xi^{1-\alpha}-1}{1-\alpha}.
\end{align*}
For $\alpha \rightarrow 1$, the $\alpha$-logarithm reduces to the usual logarithm. It is easy to see that $\log_{\alpha}(\xi)$ is strictly concave and 
\begin{align*}
    D_{\alpha}(p||q)=-\sum_j p_j \log_{\alpha}(\frac{q_j}{p_j})\ge 0,
\end{align*}
with equality if and only if $p_j=q_j$ for all $j \in I$.

Now we consider the quantum case. Given a Hilbert space $\mathcal{H}$ and two quantum density operators $\rho$ and $\sigma$ on $\mathcal{H}$, for $\alpha \in [\frac{1}{2},1) \cup (1,2]$, the Tsallis relative entropy is defined as \cite{rastegin2016quantum}
\begin{align}
    D_{\alpha}(\rho||\sigma)=\frac{\text{Tr}(\rho^{\alpha}\sigma^{1-\alpha})-1}{\alpha-1}.
\end{align}
When $\alpha \rightarrow 1$, $D_{\alpha}(\rho)$ reduces to the usual von Neumann relative entropy $S(\rho||\sigma)=\text{Tr}(\rho \log \rho-\rho \log \sigma)$. One can see that the Tsallis relative entropy is nonnegative \cite{rastegin2016quantum}: $ D_{\alpha}(\rho||\sigma)\ge 0$ with equality if and only if $\rho=\sigma$. Furthermore, the quantum Tsallis relative $\alpha$ entropy $D_{\alpha}(\rho||\sigma)$ is monotone under TPCP maps for $\alpha \in (0,1) \cup (1,2]$ \cite{rastegin2016quantum}. That is,
\begin{align}
    D_{\alpha}(\Phi(\rho)||\Phi(\sigma)) \le D_{\alpha}(\rho||\sigma),
\end{align}
for any TPCP map $\Phi$.

At the meanwhile, $D_{\alpha}(\rho||\sigma)$ is jointly convex for $\alpha \in (0,1) \cup (1,2]$ \cite{rastegin2016quantum}, i.e., if $\{p_j\}$ is a probability distribution and $\rho_j$ and $\sigma_j$ are quantum states, then
\begin{align}
    D_{\alpha}\left(\sum_j p_j \rho_j||\sum_j p_j \sigma_j \right)\le \sum_j p_j D_{\alpha}(\rho_j||\sigma_j).
\end{align}
See \cite{jenvcova2010unified} for an elegant proof of the jointly concavity and monotonicity of quantum Tsallis relative entropy.

In \cite{rastegin2016quantum}, the author introduced a coherence measure based on the quantum $\alpha$ divergence and obtained an analytic expressions as following.

For a fixed basis $\{|i\rangle\}$ of $\mathcal{H}$, $\mathcal{I}$ denotes the set of all incoherence states which are diagonal in basis $\{|i\rangle \}$. The coherence measure based on Tsallis relative entropy is defined as \cite{rastegin2016quantum}
\begin{equation} 
\begin{split}
C_{\alpha}(\rho)=&\min_{\delta \in \mathcal{I}}D_{\alpha}(\rho||\delta)\\
    =&\frac{1}{\alpha-1}\left\{\left(\sum_j \langle j|\rho^\alpha|j\rangle^{\frac{1}{\alpha}}\right)^\alpha-1\right\}.
\end{split}
\end{equation}
Although this measure enjoys some good properties such as elegant expression, the explicitly optimal incoherence state, convexity and modified strong monotonicity under incoherent operations, it violates the strong monotonicity at some special examples. For this reason, Zhao and Yu \cite{zhao2018coherence} formulated the following modified coherence measure:
\begin{equation} \label{eq1}
\begin{split}
\tilde{C}_{\alpha}(\rho)=&\min_{\delta \in \mathcal{I}}\frac{1}{\alpha-1}\left([\text{Tr}(\rho^{\alpha}\delta^{1-\alpha})]^{\frac{1}{\alpha}}-1\right)\\
    =&\frac{1}{\alpha-1}\left(\sum_j \langle j|\rho^{\alpha}|j\rangle^{\frac{1}{\alpha}}-1\right),
\end{split}
\end{equation}
for $\alpha \in (0,1) \cup (1,2]$.
This modified measure inherits some good properties of (4) and does satisfy the strong monotonicity. Moreover, the proof method provided in \cite{zhao2018coherence} is very insightful and for our use we only recall one of the observations in \cite{zhao2018coherence}:

\noindent \textbf{Observation:} If $f_{\alpha}(\rho,\sigma)=\text{Tr}(\rho^{\alpha} \sigma^{1-\alpha})$, then $f_{\alpha}(\rho,\sigma) \ge 1$ for $\alpha \in(1,2]$ and $f_{\alpha}(\rho, \sigma) \le 1$ for $\alpha \in (0,1)$ with equality if and only if $\rho=\sigma$.

The main result of this paper is to characterize the quantum correlations in bipartite system in terms of the above Tsallis type of relative entropy.

\section{Geometric measures of quantum correlations in bipartite system}
Given an Hilbert space $\mathcal{H}_A \otimes \mathcal{H}_B$. Recall that for a contractive distance $D$, the geometric measure of quantum correlations induced by $D$ is defined as
\begin{align}
    D^G(\rho)=\min_{\chi \in \mathcal{A}}D(\rho,\chi),
\end{align}
where $\rho$ is a bipartite state and $\mathcal{A}$ is the set of all classical-quantum state and for any $\chi \in \mathcal{A}$ is of the form $\chi=\sum_j q_j |j\rangle \langle j| \otimes \sigma_j$ in some basis $\{|j^A\rangle\}$.\\ 

Similarly for any bipartite state $\rho_{A B}$ on $\mathcal{H}_A \otimes \mathcal{H}_B$ we define the geometric measure of quantum correlation for $\alpha \in (0,1) \cup (1,2]$ as 
\begin{align}
    Q_{\alpha}(\rho_{AB})=\min_{\chi \in \mathcal{A}}D_{\alpha}(\rho||\chi),
\end{align}
where $\chi$ ranges over all classical-quantum state $\mathcal{A}$. For a fixed basis $\{|i^A\rangle\}$, $\chi$ is of the form
\begin{align*}
    \chi=\sum_i|i\rangle\langle i|\otimes Y_{i},
\end{align*}
where $Y_i \ge 0$ are positive operators on $\mathcal{H}_B$ and $\sum_i \text{Tr}(Y_i)=1$. In the same way, $\rho^\alpha$ can be written as 
\begin{align*}
    \rho^\alpha=\sum_{i,j}|i\rangle \langle j| \otimes R_{ij}
\end{align*}
where $R_{ij}$ are operators on $\mathcal{H}_B$ which need not all to be positive but $R_{ii}$ are exactly positive. Then 
\begin{align*}
    \text{Tr}(\rho^{\alpha}\chi^{1-\alpha})=\sum_i \text{Tr}\left(R_{ii}Y_{i}^{1-\alpha}\right).
\end{align*}
Set $X_i=R_{ii}^{1/\alpha} \ge 0$, 
\begin{align*}
    D_{\alpha}(\rho||\chi)=\frac{1}{\alpha-1}\left(\sum_i\text{Tr}(X_i^{\alpha}Y_i^{1-\alpha})-1\right),
\end{align*}
by H{\"o}lder's inequality for matrix and its equality condition \cite{larotonda2018case} and the observation in the previous section, for $0<\alpha <1$, we have
\begin{align*}
    D_{\alpha}\ge \frac{1}{\alpha-1}(\sum_i\text{Tr}(X_i)^{\alpha} \text{Tr}(Y_i)^{1-\alpha})-1),
\end{align*}
with equality if and only if $X_i= k_i Y_i$ for some $k_i \ge 0$. Then if we set $\text{Tr}(X_i)=r_i$ and $\text{Tr}(Y_i)=s_i$, we have
\begin{equation*}
\begin{split}
 &D_{\alpha}(\rho||\chi)\\
    \ge& \frac{1}{\alpha-1}\left(\sum_i r_i^{\alpha}s_i^{1-\alpha}-1\right)\\
    =&N^{\alpha}D_{\alpha}\left(\frac{r_i}{N}||s_i\right)+\frac{N^{\alpha}-1}{\alpha-1}\\
    \ge&\frac{N^{\alpha}-1}{\alpha-1},
\end{split}
\end{equation*}
where $N=\sum_i \text{Tr}(X_i)=\sum_i r_i$ and the second inequality is due to $D_{\alpha}\left(\frac{r_i}{N}||s_i\right) \ge 0$ with equality if and only if  $\frac{r_i}{N}=s_i$. The first inequality with equality if and only if $X_i=k_i Y_i$ for some $k_i \ge 0$, which means that $r_i= N s_i=k_i s_i$. This is equivalently to say that $D_{\alpha}(\rho||\chi)=\frac{N^{\alpha}-1}{\alpha-1}$ if and only if $X_i=\frac{1}{N}Y_i$, thus we can conclude that the minimum can be obtained if $Y_i=\frac{1}{N}X_i$.
Then for $Q_{\alpha}(\rho_{AB})$, we obtain the analytic expression as
\begin{align}
    Q_{\alpha}(\rho_{AB})=\min_{\chi \in \mathcal{A}}D_{\alpha}(\rho_{AB}||\chi)=\min_{\{|i^A\rangle\}}\frac{N^{\alpha}-1}{\alpha-1},
\end{align}
where $N=\sum_i \text{Tr}(X_i)$ and $X_i=\langle i|\rho^{\alpha}|i\rangle^{1/\alpha}$. Furthermore, the optimal classical-quantum state $\chi$ is of the form $\chi=\sum_i |i\rangle \langle i| \otimes Y_i$ where $Y_i=\frac{1}{N}\langle i|\rho^{\alpha}|i\rangle^{1/\alpha}$.

Note that the above derivation is only focus on $0 < \alpha <1$, but for $ 1<\alpha <2$, by the reverse H"older's inequality and its equality condition and the observation in preliminaries, our conclusion preserves unchanged. Moreover, if we set $\alpha=\frac{1}{2}$, we obtained the correlation measure based on the Hellinger distance, see \cite{roga2016geometric} for more details. Thus our work generalized the results in \cite{roga2016geometric} $\alpha=\frac{1}{2}$ to $ \alpha \in (0,1) \cup (1,2]$ case.

In the next we show that $Q_{\alpha}(\rho_{AB})$ satisfies some properties:
\begin{enumerate}
    \item $Q_{\alpha}(\rho_{AB})\ge 0$ with $Q_{\alpha}(\rho)=0$ if and only if $\rho$ is a classical-quantum state;
    \item $Q_{\alpha}(\rho_{AB})=Q_{\alpha}(U_A \otimes U_B \; \rho_{AB} \;U_A^{\dagger}\otimes U_B^{\dagger})$;
    \item $Q_{\alpha}(\rho_{AB}) \ge Q_{\alpha}(\Phi_B(\rho_{AB}))$ for local TPCP map $\Phi_B$ on system $B$.
\end{enumerate}

For property 1, $Q_{\alpha}(\rho_{AB}) \ge 0$ by definition. If $\rho_{AB}$ is a classical-quantum state, from the deduction of equality (8) we can find an optimal state in $\mathcal{A}$. On the other hand, if $Q_{\alpha}(\rho_{AB})=0$, since $Q=\min_{\chi \in \mathcal{A}}D_{\alpha}(\rho_{AB})$, there exists a state $\chi_0 \in \mathcal{A}$ such that $D_{\alpha}(\rho_{AB}||\chi_0)=0$. The Tsallis relative entropy $D_{\alpha}(\rho||\sigma)=0$ if and only if $\rho=\sigma$ for all $\rho$ and $\sigma$, we can conclude that $D_{\alpha}(\rho_{AB}||\chi_0)=0$ if and only $\rho_{AB}=\chi_0 \in \mathcal{A}$.  Property 3 can be obtained from the monotonicity of $D_{\alpha}(\rho||\sigma)$, so we only need to prove property 2. Since 
\begin{align*}
    D_{\alpha}(U_A \otimes U_B \;\rho_{AB}\; U_A^{\dagger}\otimes U_B^{\dagger})=D_{\alpha}(\rho||\chi^{\prime}),
\end{align*}
where $\chi^{\prime}=U_A^{\dagger}\otimes U_B^{\dagger} \;\chi \;U_A \otimes U_B$ is also a classical-quantum state in another local basis $\{|j^A\rangle\}$. After the minimization over all local basis $\{|i^A\rangle\}$, the results follows.

Thus we formulate a new class of bona fide geometric measures of quantum correlations in bipartite system based on Tsallis relative entropy.

\section{Correlation measures based on the modified quantum Tsallis relative entropy}
Inspired by the work in \cite{zhao2018coherence}, in this section, we show that the modified quantum Tsallis relative entropy can also be used to quantify quantum correlations.
Denote $f(\rho,\sigma)=\text{Tr}(\rho^\alpha \sigma^{1-\alpha})$, a new coherence measure can be defined as \cite{zhao2018coherence}
\begin{align}
    C_{\alpha}(\rho)=\min_{\delta \in \mathcal{I}}\frac{f^{1/\alpha}(\rho,\delta)-1}{\alpha-1}.
\end{align}
Similarly, we can define a correlation measure as
\begin{align}
    \tilde{Q}_{\alpha}(\rho_{AB})=\min_{\chi \in \mathcal{A}}\frac{f^{1/\alpha}(\rho,\chi)-1}{\alpha-1},
\end{align}
by virtue of the same proof method in section \rom{3} and the fact of the monotonicity of function $g(x)=x^{t}$ for any fixed $t >0$, we have
\begin{align}
    \tilde{Q}_{\alpha}(\rho_{AB})=\min_{|i^A\rangle}\frac{N-1}{\alpha-1}.
\end{align}
Moreover, when $\alpha=\frac{1}{2}$, our correlation measure is just the correlation measure defined in \cite{yu2014quantum} which is induced by skew information, up to a factor 2. By the way, if $\rho_{AB}$ is a pure bipartite state $\rho_{AB}=|\psi_{AB}\rangle \langle \psi_{AB}|$ where $|\psi_{AB}\rangle=\sum_j \sqrt{\lambda_j}|j_A\rangle|j_B\rangle$ is its Schmidt decomposition, by simple calculation,
\begin{align*}
    \tilde{Q}_{\alpha}(|\psi_{AB}\rangle)=\frac{\sum_j\lambda_j^{1/\alpha}-1}{\alpha-1},
\end{align*}
which is a measure of entanglement in terms of Tsallis entropy \cite{luo2016general}, up to a constant factor.

\section{Some Illuminated examples}
In this section, we present the analytic form of the geometric measure for Werner state and isotropic state in arbitrary dimensions.

\textit{Example 1.} The $d \otimes d$ Werner state is 
\begin{equation}
    \rho_W=\frac{d-x}{d^3-d}I_d \otimes I_d + \frac{d x-3}{d^3-d}V, \quad x \in [-1,1],
\end{equation}
with $V=\sum_{ij}|ij\rangle \langle  ji|$ the swap operator. From the description previous, the key point is the calculation of $N=\sum_i \text{Tr} \langle i|\rho^{\alpha}|i\rangle$. In this case, 
\begin{equation}
    N_W=\frac{1+x}{d+1}+(d-1)\left\{\frac{1}{2}\left(\frac{1+x}{d+1}\right)^{\alpha} + \frac{1}{2}\left(\frac{1-x}{d-1}\right)^{\alpha}\right\}^{1/\alpha}.
\end{equation}
$Q_\alpha(\rho_W)=0$ when $x=\frac{1}{d}$. Also
if $\alpha=\frac{1}{2}$, our result coincides with \cite{yu2014quantum}\cite{sun2017quantum}.

\textit{Example 2.} The $d \otimes d$ isotropic state is  
\begin{equation}
\rho_I= \frac{1-x}{d^2-1}I_d \otimes I_d + \frac{d^2 x-1}{d^2-1}|\Phi\rangle \langle \Phi|, \quad x\in [0,1],
\end{equation}
where $|\Phi\rangle=\frac{1}{\sqrt{d}}\sum_k |kk\rangle$. The key quantity $N$ can be obtained as
\begin{equation}
    N_I=\frac{d(1-x)}{d^2-1}+\left\{(m-1)\left(\frac{1-x}{d^2-1}\right)^{\alpha}+ x^{\alpha}\right\}^{1/\alpha}.
\end{equation}
And when $x=\frac{1}{d^2}$, $Q_{\alpha}(\rho_I)=0$.

\section{Quantum coherence measure in L"uders measurement picture}
Recently, the coherence measures are generalized to L"uders measurement picture as a partial coherence quantification \cite{sun2017quantum} \cite{luo2017quantum} \cite{luo2017partial} \cite{luo2018coherence}. In this section we show that the coherence measure in terms of quantum Tsallis relative entropy can be also generalized into this setting.

For a Hilbert space $\mathcal{H}$ and a fixed l"uders measurement $\{\Pi_j, j=1,\cdots, m\}$, the incoherent states with respect to this L"uders measurement can be written as
\begin{align}
    \mathcal{I}_L=\{\delta\; | \; \sum_j \Pi_j \,\delta \,\Pi_j =\delta\}.
\end{align}
For any density operator $\rho$ on $\mathcal{H}$ and any $\alpha \in (0,1) \cup (1,2]$, we define the partial coherence measure as
\begin{align}
    C_{L,\alpha}(\rho)=\min_{\delta \in \mathcal{I}_L}D_{\alpha}(\rho||\delta).
\end{align}
$\rho^\alpha$ can be written as 
\begin{align*}
    \rho^{\alpha}=\begin{pmatrix}
                   R_{11} & \cdots & R_{1m}\\
                   \vdots & \vdots  & \vdots\\
                   R_{m1} & \cdots  & R_{mm}
        \end{pmatrix},
\end{align*}
where the partition is in conformable with the L{\"u}ders measurement $\{\Pi_j\}$, and $R_{ii} $ are positive. Similarly, for any $\delta \in \mathcal{I}_L$, 
\begin{align*}
    \delta=\begin{pmatrix}
    Y_1 & 0 & \cdots & 0\\
    0 & Y_2 & \cdots & 0\\
    \vdots &\vdots & \vdots & \vdots\\
    0 & 0 & \cdots & Y_m
    \end{pmatrix}.
\end{align*}
Set $X_i=R_{ii}^{1/\alpha}$, and by similar calculation in previous section, we have
\begin{align}
    C_{L,\alpha}(\rho)=\frac{N^\alpha-1}{\alpha-1},
\end{align}
where $N=\sum_j \text{Tr}\{(\Pi_j \rho^{\alpha}\Pi_j)^{\frac{1}{\alpha}}\}$.

Note that recently Rastegin \cite{rastegin2018coherence} had calculated the same quantity as ours, but we think our results is more precise, since we consider the general incoherent state, or equivalently, the block diagonal states with respect to the given L{\"u}ders measurement.

Furthermore, for $\alpha \in (0,1) \cup (1,2]$, if we define 
\begin{align}
    \tilde{C}_{L,\alpha}(\rho)=\min_{\delta \in \mathcal{I}_L}\tilde{D}_{\alpha}(\rho||\delta),
\end{align}
we can obtain the analytic expression:
\begin{align}
    \tilde{C}_{L,\alpha}(\rho)=\frac{N-1}{\alpha-1},
\end{align}
where $N=\sum_j \text{Tr}\{(\Pi_j \rho^{\alpha}\Pi_j)^{\frac{1}{\alpha}}\}$. And the optimal incoherent state is 
\begin{align*}
    \delta^*_L=\frac{1}{N}\sum_j \left(\Pi_j \rho^\alpha \Pi_j \right)^{1/\alpha}. 
\end{align*}
What is more, if we set $\alpha=\frac{1}{2}$, we reproduce the quantum uncertainty measure formulated in \cite{luo2017quantum}, up to a factor 2.\\

\section{Concluding remarks}
In this paper we formulated a class of new quantum correlation measures which based on quantum Tsallis relative entropy and its modified version, which has been shown not only enjoy elegant analytic expressions, but also contains many novel known quantum correlation measures as special cases. It is also can be viewed as an one-parameter extension of the von Neumann relative entropy correlation measure. The future work is to find some applications and operational interpretations in quantum information processing tasks.

\section{Acknowledgments}
The author is very grateful to professor Runyao Duan and professor Shunlong Luo for insightful discussions.

\bibliographystyle{unsrt}
\bibliography{bib.bib}

\end{document}